\documentclass{article}
\usepackage{amsfonts}
\usepackage{amsmath}
\usepackage{epsfig}

\begin{document}

\title{Proposal For Testing Bell's Conjecture}
\author{Luiz Carlos Ryff \\
\textit{Instituto de F\'{\i}sica, Universidade Federal do Rio de Janeiro, }\\
\textit{Caixa Postal 68528, 21941-972 RJ, Brazil}\\
{\small e-mail: ryff@if.ufrj.br}}
\maketitle

\begin{abstract}
A simple nonlocal mechanism for Einstein-Podolsky-Rosen correlations
inspired by Bell's conjecture (according to which "behind the scenes
something is going faster than light") is suggested, and an experimental
test is proposed.\smallskip

\end{abstract}

\bigskip

\bigskip

\textbf{I. INTRODUCTION\bigskip }

\bigskip

EPR correlations \textrm{[1] }are important not only due to their potential
applications in quantum computing and cryptography \textrm{[2]}, but also
from the standpoint of the foundations of quantum theory \textrm{[3]}.
Physicists are divided on this mind-boggling phenomenon that suggests a
nonlocal connection between entangled particles. Some, based on experimental
loopholes, believe --- as Einstein did --- that a local theory will complete
quantum mechanics $\mathrm{[4]}$, while others consider that realism has to
be revised $\mathrm{[}$\textrm{5];} a few, following John Bell, think that
"behind the scenes something is going faster than light" $\mathrm{[6];}$
many, agreeing with Asher Peres, advocate the straightforward application of
the quantum mechanical formalism with no need for any metaphysical
interpretation $\mathrm{[}$\textrm{7}$\mathrm{]}$, and a not inconsiderable
fraction probably simply ignores the problem \textrm{[8]}. Among those
showing some sympathy for a nonlocal interaction approach are Gisin $\mathrm{%
[9]}$, and Leggett $\mathrm{[}$\textrm{10}$\mathrm{]}$. Bell and Bohm were
very explicit about the possible existence of some kind of superluminal
interaction $\mathrm{[}$\textrm{11}$\mathrm{]}$, and the former has made it
very clear that the assumption of a preferred frame of reference is not
necessarily inconsistent with Lorentz transformations $\mathrm{[12]}$. In
this paper, inspired by the ideas of Bell, I will suggest a simple mechanism
that might be behind these strange correlations and show how it could be
checked.

In an attempt to investigate Bell's conjecture, I will assume (considering a
two-photon entangled state) that there must be some sort of wave, that I
will call a Bell-wave (or B-wave, for short), that propagates from the first
detected photon (in the preferred frame) to the second, "forcing" it into a
well-defined state (although the natural candidate responsible for the
unleashing of this process is the detection of the first photon, other
possibilities will be discussed in \textrm{Sec. III}). I will also assume,
as a working hypothesis, that this B-wave can interact with physical
objects, such as beam splitters, mirrors, phase shifters, polarizers and so
forth, such that a nonstatic experiment might lead to totally different
results from a static experiment. To develop this idea further, it is
necessary to examine two important experimental tests of Bell's inequalities
that used time varying analyzers.

The first one, uses acousto-optical switches $\mathrm{[13]}$. Entangled
photons, $\nu _{1}$ and $\nu _{2}$, from a source $S$ can reach polarizers
\textrm{I} and \textrm{II}, when the switches are in the transmission mode,
and polarizers \textrm{I' }and \textrm{II'}, when they are in the reflection
mode. The second uses Pockels cells \textrm{(}$C_{1}$ \textrm{and} $C_{2}$
\textrm{in} \textrm{Fig. 1)} and two-channel polarizers, one in each arm of
the experimental apparatus \textrm{[14]}. But now, differently from the
first experiment, the switches work randomically as opposed to periodically.
When the cell is activated this is equivalent to a rotation of the
corresponding polarizer. The motivation for this type of experiment is to
have the settings change during the time of flight of the particles so that
no exchange of signals with velocity v$\ \leq $ $\ $c$\ \ $could be
responsible for the violations of Bell's inequalities$\ \mathrm{[15]}$. The
violation of a CHSH inequality \textrm{[16] }indeed discarded this
possibility. The motivation behind the experiment I want to discuss is to
have one of the switches in one mode when the photon passes through it
coming from the source, and in another mode when the B-wave passes through
it in the opposite direction coming from the detector. For instance, if $%
C_{1}$ is inactivated (activated) when $\nu _{1}$ impinges on it, it has to
be activated (inactivated) when $\nu _{1}$ is detected (I am assuming that
"detection"\textit{\ }or, strictly speaking, "photon absorption", is an
objective fact, independently of it being registered or not). But the
changing of mode and the detection are space-like events, therefore their
relative position in time depends on the Lorentz frame we use to observe
them. It is easy to see $\mathrm{(Sec.}$ \textrm{II)} that there are an
infinite number of frames in which the above condition is never satisfied.
From Bell's nonlocal point of view, there must be a preferred frame in which
one of these events \textit{really }occurs before the other. But we do not
know how to determine this frame. However, if the geometry of the experiment
is changed so that the events become separated by a time-like interval,
their relative position in time will be the same in all Lorentz frames. Now,
in principle, they could be causally connected. Therefore, it is possible to
know which event really occurred first. In the experiment represented in
\textrm{Fig. 2} the light path between $C_{1}$ $(C_{2})$ and $D_{1}$ $%
(D_{2}) $ and $D_{1^{^{\prime }}}$ $(D_{2^{^{\prime }}})$ has been made
longer by means of a detour, so that, in any Lorentz frame, when $\nu _{1}$ $%
(\nu _{2}) $ is detected $C_{1}$ $(C_{2})$ is already in a different state
from that which it was in when $\nu _{1}$ $(\nu _{2})$ passed through it
\textrm{(Sec. III)} (naturally, now the activation of the cells cannot be
randomic). If $\nu _{1}$ $(\nu _{2})$ is detected first in the preferred
frame, the B-wave goes from $\nu _{1}$ $(\nu _{2})$ to $\nu _{2}$ $(\nu
_{1}) $. Alternatively, we could have an asymmetric arrangement, in which
the light path between $S$ and $D_{2}$ $(D_{1})$ and $D_{2^{^{\prime }}}$ $%
(D_{1^{^{\prime }}})$ would be made longer, so that $\nu _{2}$ $(\nu _{1})$
would always impinge on the polarizer after the detection of $\nu _{1}(\nu
_{2})$ \textrm{(Fig. 3)}. In this case, the B-wave would always go from $\nu
_{1}$ $(\nu _{2})$ to $\nu _{2}$ $(\nu _{1})$. If the mechanism suggested
here is correct, we will no longer observe the correlations predicted by
quantum mechanics.

\bigskip

\bigskip

\textbf{II. NONSTATIC\ TESTS\ OF\ BELL'S\ INEQUALITIES\ SEEN\ FROM\ MOVING
FRAMES}

\bigskip

\bigskip

I will discuss the experiment of \textrm{ref. 14}, though the same
argumentation can be used for the experiment of \textrm{ref. 13}.\ We can
consider a specific situation. Let $t_{0}=0$ be the instant at which $\nu
_{1}$ and $\nu _{2}$ are emitted in laboratory frame $L$ $\mathrm{(Fig.}$
\textrm{1)}, $t_{C_{1}}(t_{C_{2}})$ the instant at which $C_{1}(C_{2})$
completely changes (after the passage of the photon) from the i-mode
(inactivated), to the a-mode (activated), and $\overline{t}$ the instant at
which $\nu _{1}$ and $\nu _{2}$ are absorbed at detectors $D_{1}$ and $D_{2}$
(a similar reasoning would be valid for $D_{1^{^{\prime }}}$ and $%
D_{2^{^{\prime }}}$). I will assume that when the photons are detected the
switches are already in the a-mode. But, as can easily be seen, there are an
infinite number of Lorentz frames in which they are still in the i-mode. For
example, let us observe the experiment from a frame $L^{\prime }$

\bigskip
\begin{figure}[tbp]
\centering
\par
\hspace*{+0.3in}
\par
\includegraphics[height= 2.5
cm,width=12.0cm]{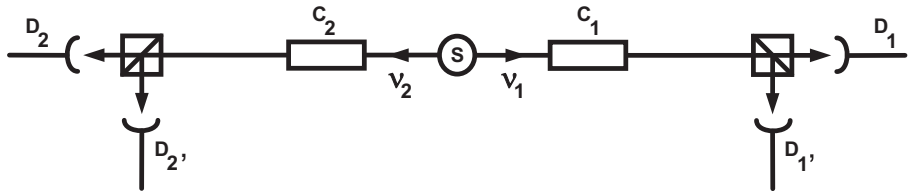} \caption{ A nonstatic test of Bell's
inequalities.}
\end{figure}


\bigskip

\noindent moving in the same direction as $\nu _{1}$ with velocity $v$. Let $%
t_{0}^{\prime }=0$ be the instant at which the photons are emitted in $%
L^{\prime }$. Hence, the instants at which $C_{1}$ and $C_{2}$ change
completely are given by
\begin{equation}
t_{C_{1}}^{\prime }=\gamma \left( t_{C_{1}}-\frac{v}{c^{2}}x\right)  \tag{1}
\end{equation}%
and
\begin{equation}
t_{C_{2}}^{\prime }=\gamma \left( t_{C_{2}}+\frac{v}{c^{2}}x\right) ,
\tag{2}
\end{equation}%
where $\gamma =1/\sqrt{1-v^{2}/c^{2}},$ $x$ $(-x)$ is the distance from the
source $S$ to $C_{1}$ $(C_{2})$, and I am assuming that $L$ and $L^{\prime }$
are in the standard configuration. For the instants at which $\nu _{1}$ and $%
\nu _{2}$ are absorbed we obtain
\begin{equation}
\overline{t}_{1}^{\prime }=\gamma \left( \overline{t}-\frac{v}{c^{2}}%
\overline{x}\right)  \tag{3}
\end{equation}%
and
\begin{equation}
\overline{t}_{2}^{\prime }=\gamma \left( \overline{t}+\frac{v}{c^{2}}%
\overline{x}\right) ,  \tag{4}
\end{equation}%
where $\overline{x}$ $(-\overline{x})$ is the distance from $S$ to $D_{1}$ ($%
D_{2}$). To have $\nu _{1}$ detected before $C_{1}$ changes to the a-mode,
the condition%
\begin{equation}
\overline{t}_{1}^{\prime }<t_{C_{1}}^{\prime }  \tag{5}
\end{equation}%
must be fulfilled, which leads, using $(1)$ and $(3)$, to
\begin{equation}
v>c\left( \frac{\overline{t}-t_{C_{1}}}{t_{f}}\right) ,  \tag{6}
\end{equation}%
where $t_{f}=(\overline{x}-x)/c$ is the time the photon takes to go from $%
C_{1}$ to $D_{1}$. In a frame moving with a velocity that satisfies
condition $(6)$ $\nu _{1}$ will be detected before $C_{1}$ changes its
state. For $\nu _{1}$ to be detected before $C_{2}$ changes to the a-mode,
the following condition must be fulfilled:
\begin{equation}
\overline{t}_{1}^{\prime }<t_{C_{2}}^{\prime }.  \tag{7}
\end{equation}%
Since $t_{C_{2}}\approx t_{C_{1}}$, we see from $(1)$ and $(2)$ that $(5)$
implies $(7)$. (It is easy to see that $(7)$ leads to $v>c\times \left(
\overline{t}-t_{C_{2}}\right) /T_{f}$, where $T_{f}=(\overline{x}+x)/c$%
.)\bigskip \bigskip

\bigskip \textbf{III. A PROPOSED EXPERIMENT\bigskip }

\bigskip

The experiment represented in \textrm{Fig. 2 }is different from the
experiment of \textrm{ref. 14} in two respects: there is a detour in each
photon path and the switches work periodically. Initially, I will consider
the following situation. Photon $\nu _{1}$ (the same reasoning is valid for $%
\nu _{2}$) reaches $C_{1}$ when it has just entered the i(a)-mode. Let $%
\Delta t$ be the duration of each mode, and $\overline{t}_{f}$ the time
taken by $\nu _{1}$ to go from $C_{1}$ to $D_{1}$. The condition that the
detection of $\nu _{1}$ \textit{really }occurs after $C_{1}$ has already
completely changed to the a(i)-mode can be expressed as \textrm{[17]}
\begin{equation}
\Delta t+t_{f}<\overline{t}_{f},  \tag{8}
\end{equation}%
since, after the passage of the photon, $C_{1}$ takes a time interval of $%
\Delta t$ to change completely from the i(a)-mode to the a(i)-mode, and an
imaginary light beam sent from $C_{1}$ at this moment would take time $t_{f}$
to reach $D_{1}$ following a straight line. According to $(8)$, this light
beam would reach $D_{1}$ before $\nu _{1}$. But,
\begin{equation}
\overline{t}_{f}=t_{f}+\frac{2y}{c},  \tag{9}
\end{equation}%
where $y$ is the height of the detour. Using $(8)$ and $(9)$, we obtain the
following condition that our experiment has to fulfil:
\begin{equation}
y>\frac{c\Delta t}{2}.  \tag{10}
\end{equation}%
\bigskip

It is also necessary that when $\nu _{1}$ is detected $C_{1}$ has not yet
returned to the i(a)-mode. The condition that the detection of $\nu _{1}$
really occurs before $C_{1}$ has already returned to the i(a)-mode can be
expressed as
\begin{equation}
\overline{t}_{f}+t_{f}<2\Delta t.  \tag{11}
\end{equation}%
Using $(9)$ and $(11)$, we also obtain
\begin{equation}
y<c\Delta t-ct_{f}.  \tag{12}
\end{equation}%
Then, $(10)$ and $(12)$ lead to
\begin{equation}
t_{f}<\frac{\Delta t}{2}.  \tag{13}
\end{equation}%
But, $t_{f}=(\overline{x}-x)/c$, where $\overline{x}-x$ is the distance
between $C_{1}$ and $D_{1}$. So, we must also have
\begin{equation}
\overline{x}-x<\frac{c\Delta t}{2}.  \tag{14}
\end{equation}

\bigskip

\begin{figure}[tbp]
\centering
\par
\hspace*{+0.3in}
\par
\includegraphics[height= 2.5
cm,width=12.0cm]{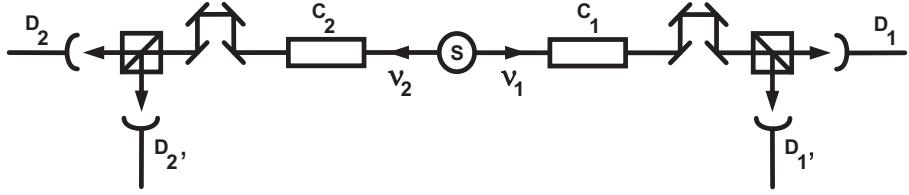} \caption{The proposed experiment. (The
relative sizes do not correspond to that of an actual experiment.)}
\end{figure}


\ \ \ \ \ \

Let us now consider the situation in which $\nu _{1}$ reaches $C_{1}$ just
before it leaves the i(a)-mode. Now, instead of $(8)$ we have $t_{f}<%
\overline{t}_{f}$, which is always satisfied. We see that, if condition $%
(10) $ is fulfilled, all photons, independently of the instant they reach $%
C_{1}$, will be detected after the change from the i(a)-mode to the
a(i)-mode. On the other hand, instead of $(11)$ we have $\overline{t}%
_{f}+t_{f}<\Delta t$. Then using $(9)$, we obtain $2y<c\Delta t-2ct_{f}$,
which is inconsistent with $(10)$. A possible way to overcome this
difficulty is to try to synchronize the emission of the photons with the
activation and inactivation of the switches, such that the photons will
always reach the switches when they have just entered a mode. In this case,
conditions $(10)$, $(12)$ and $(14)$ will be satisfied.

Another way to overcome the difficulty is based on a compromise solution.
For example, let us consider the photons $\nu _{1}$\ that reach $C_{1}$ when
it has already spent a time equal to $9\Delta t/10$ in the i(a)-mode. Then,
instead of $(8)$, we must have
\begin{equation}
\frac{\Delta t}{10}+t_{f}<\overline{t}_{f},  \tag{15}
\end{equation}%
which leads, using $(9)$, to
\begin{equation}
y>\frac{c\Delta t}{20}.  \tag{16}
\end{equation}%
Therefore, if $(10)$ is satisfied, this is also true for $(16)$. On the
other hand, instead of $(11)$ we must have
\begin{equation}
\overline{t}_{f}+t_{f}<\frac{\Delta t}{10}+\Delta t,  \tag{17}
\end{equation}%
which leads, using $(9)$, to
\begin{equation}
y<\frac{11c\Delta t}{20}-ct_{f}.  \tag{18}
\end{equation}%
Hence, using $(10)$, we obtain
\begin{equation}
t_{f}<\frac{\Delta t}{20}  \tag{19}
\end{equation}%
and
\begin{equation}
\overline{x}-x<\frac{c\Delta t}{20},  \tag{20}
\end{equation}%
which is more restrictive than $(14)$. If conditions $(10)$, $(18)$ and $%
(20) $ are fulfilled, 90\% of the photons will effectively\textit{\ }reach
the switches when they are in the i(a)-mode, and will be detected when they
are in the a(i)-mode. Using logic circuits \textrm{[14]}, the other 10\% of
the events, that correspond to the photons that reach the switch when it has
already spent a time greater than $9\Delta t/10$ in the i(a)-mode, can be
disregarded.

Although I have only discussed detections at $D_{1}$, it is easy to see
that, if the distance from $C_{1}$ to $D_{1^{^{\prime }}}$ is smaller than
the distance from $C_{1}$ to $D_{1}$, as in the scheme represented on
\textrm{Fig. 2}, then the deduced conditions contain those for detections at
$D_{1^{^{\prime }}}$. Detections "without a click" \textrm{[18]} are also
included. With respect to this point, it is interesting to observe that, if $%
D_{1^{^{\prime }}}$ is in the "right" position, the condition $(20)$ for $%
D_{1}$ does not need to be fulfilled, since (considering the situation in
which both photons are detected) the lack of detection at $D_{1^{^{\prime
}}} $ is equivalent to a detection at $D_{1}$ in the "right" position (in
ideal situations, this is true even if $D_{1}$ has been removed!) It is also
easy to see that the assumption that the B-wave is triggered by the
splitting of the photon at the polarizer into a "photonic" and an "empty"
wave --- which is consistent with the pilot wave interpretation \textrm{[19]
}--- leads to conditions that are also contained in those that have been
obtained. In the event of corroboration of Bell's conjecture, the proposed
experiment can be modified in order to determine where and when the collapse
of the state vector (the triggering of the B-wave, in the present case)
takes place: in the polarizer, when the photon is split, or in the detector,
when it is annihilated (or when it is not, if we have detection without a
click). For this, we can place the detours between the polarizers and the
detectors. If the results agree (disagree) with the predictions of quantum
mechanics, the first (second) possibility is the correct one.

In conclusion, the strong correlations displayed by twin photons, which
result from entanglement, suggests, as strangely as this may sound, a
nonlocal connection between these photons. This point has been emphasized by
Bell, according to whom "behind the scenes something is going faster than
light". A simple mechanism for this process, suggested here, can be
experimentally tested by means of a \ straightforward modification of two
experiments already performed to test the CHSH inequality. Although only
qualitative predictions are being made, this kind of approach is not foreign
to physics, and is in agreement with its investigative nature, the best
example being the discovery of the law of electromagnetic induction by
Faraday \textrm{[20]}. However, for the sake of completeness, a simple
nonlocal model that yields quantitative predictions is discussed in the
\textrm{Appendix}.

\bigskip \bigskip

\bigskip

\textbf{APPENDIX: A SIMPLE NONLOCAL MODEL}

\bigskip

\bigskip

The following nonlocal model bears some resemblance to the advanced wave
interpretation \textrm{[21]}, though now the detector that has \textit{really%
} "clicked" first effectively emits a B-wave. It is easier to discuss the
experiment represented in \textrm{Fig. 3}. Now $\nu _{1}$ is always detected
before $\nu _{2}$, and the conditions obtained in \textrm{Sec. III }are
valid for it. As an example, I will consider a specific situation; other
situations can be treated in a similar way. Both polarizers are oriented
vertically, and the Pockels cell $C_{1}$, when activated, rotates the
polarization of light by an angle $-\theta $. Let us represent the
two-photon entangled state emitted by $S$ as $\mid \Psi \rangle =1/\sqrt{2}%
\left( \mid V\rangle _{2}\mid H\rangle _{1}-\mid H\rangle _{2}\mid V\rangle
_{1}\right) $ \textrm{[22]}. To be in agreement with the quantum mechanical
predictions for the static case (i.e., when $C_{1}$ is permanently \ either
activated or inactivated) I will assume that the B-wave emitted by $%
D_{1}(D_{1^{\prime }})$ emerges from the polarizer in "state" $\mid V\rangle
_{B}\equiv \mid 0\rangle _{B}$ $(\mid H\rangle _{B}\equiv \mid \frac{\pi }{2}%
\rangle _{B})$ \textrm{[23]}; then, passing through $C_{1}$, it emerges
either in the same state (if $C_{1}$ is inactivated) or in state $\mid
\theta \rangle _{B}(\mid \frac{\pi }{2}+\theta \rangle _{B})$ (if $C_{1}$ is
activated). In this respect, the B-wave behaves like an ordinary
electromagnetic wave. It then goes into the source and emerges in an
orthogonal polarization state \textrm{[24]}. This is the state into which $%
\nu _{2}$ will be forced, if no other optical devices are on the path of the
B-wave. I will further assume that the these rules are still valid in the
nonstatic case.\newpage

\bigskip

\begin{figure}[tbp]
\centering
\par
\hspace*{+0.3in}
\par
\includegraphics[height= 3.5
cm,width=12.0cm]{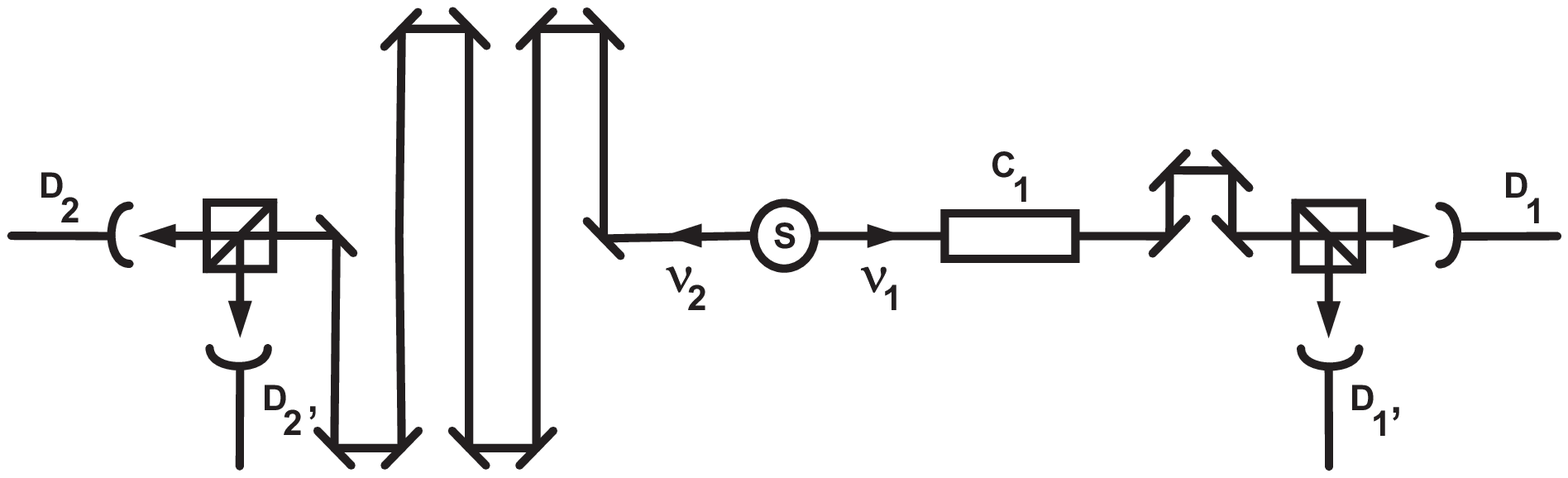} \caption{A variant of the proposed
experiment. (The relative sizes do} {\small not correspond to that
of an actual experiment.)}
\end{figure}


\bigskip

Let us then examine the situation in which $\nu _{1}$ passes through $C_{1}$
when it is inactivated. From the standpoint of quantum mechanics the
detection probabilities do not depend on which photon is detected first. The
probability of having $\nu _{2}$ detected in the state $\mid V\rangle $ is $%
1/2$, in which case $\nu _{1}$ will be forced into the state $\mid H\rangle $%
, and the probability of having both photons transmitted will be $%
(P_{21})_{Q}=0$. On the other hand, from the nonlocal point of view
introduced here, $C_{1}$ will be activated for the B-wave, since $\nu _{1}$
is effectively detected first, and we will have: $\mid V\rangle
_{B}\rightarrow \mid \theta \rangle _{B}$, after $C_{1}$, and $\mid \theta
\rangle _{B}\rightarrow \mid \frac{\pi }{2}+\theta \rangle $, after $S$,
which leads to $(P_{21})_{B}=\frac{1}{2}\sin ^{2}\theta $.


\begin{thebibliography}{99}
\bibitem{8} Naturally, this is not intended to be an accurate and complete
description of the different points of view related to the conceptual
consequences of EPR correlations.
\bibitem{17} From an aether point of view, $(8)$ is only a sufficient
condition. It expresses the fact that $C_{1}$ changes from the i(a)-mode to
the a(i)-mode before the detection of $\nu _{1}$, and that these two events
are separated by a time-like interval.
\bibitem{23} If the orientation of the polarizer is parallel to an arbitrary
$\mathbf{a}$, for example, the B-wave emerges in state $\mid a_{\parallel
}\rangle _{B}(\mid a_{\perp }\rangle _{B})$.
\bibitem{24} If the state emitted by $S$ were $\mid \Psi \rangle =1/\sqrt{2}%
(\mid V\rangle _{2}\mid V\rangle _{1}+\mid H\rangle _{2}\mid H\rangle _{1})$%
, the B-wave would emerge from the source with the polarization unchanged.
\end{thebibliography}
\end{document}